\documentstyle{article}
\begin{document}

\title{U(1) Gauge Theory as Quantum Hydrodynamics}
\author{ Girish S. Setlur \\ The Institute of Mathematical Sciences
 \\ Taramani, Chennai 600113 }
 
\maketitle                   

\begin{abstract}
 It is shown that gauge theories are most naturally studied via a polar
 decomposition of the field variable. Gauge transformations 
 may be viewed as those that leave the density invariant but
 change the phase variable by 
 additive amounts. The path integral approach is used to compute
 the partition function. When gauge fields are included, the constraint
 brought about by gauge invariance simply means an appropriate
 linear combination of the gradients of the 
 phase variable and the gauge field is 
 invariant. No gauge fixing is needed in this approach that is
 closest to the spirit of the gauge principle. 
 We derive an exact formula for the
 condensate fraction and in case it is zero, an exact
 formula for the anomalous exponent. We also derive a formula for the vortex
 strength which involves computing radiation corrections.
\end{abstract}

\section{Introduction}

 The density phase transformation for bosons is quite well-known
 to those who work in the fields of
 superfluidity\cite{fluid1} \cite{fluid2}
 and other related areas such as superconductivity and possibly
 even quantum optics.  
 Jackiw and collaborators\cite{Jackiw} have recently introduced some
 of these ideas independently in the context of relativisitic
 quarks. Since the subject of bosons is vast, 
 we refer the reader to the review by Ceperley\cite{Cep} for
 further references. However a few basic works do deserve mention.
 First there is the work by Penrose and Onsager\cite{Pen}   
 on estimating the superfluid fraction in $ He^{4} $.
 Then there is the Hohenberg-Mermin-Wagner theorem\cite{Hohen}
 that precludes long-range order with a broken continuous symmetry 
 in two dimensions. 
 Finally, there is work by Ceperley\cite{Cep} using Quantum Monte Carlo.

 The main point of this article is to highlight the usefullness 
 of the hydrodynamical approach in tackling gauge theories. We argue that this
 approach is the natural way in which to study gauge theories. 
 Although by itself none of the results in this article 
 are particularly new, the novel aspect is the manner in which the theory
 is formulated that renders it susceptible to generalization 
 in future publications to fermions coupled to gauge fields and later
 on to relativistic gauge theories
 and finally to nonabelian relativistic gauge
 theories. Much of this agenda has already been completed
 by the high energy community recently, specifically, the work
 of Jackiw and Polychronakos\cite{Jackiw} in introducing anticommuting
 Grassmann variables into a fluid dynamical description will be very relevent
 to our future work. But our thrust is to compute microscopic correlation
 functions rather than just develop formalism. 

 One concrete result in this article is an exact formula for
 the condensate fraction. Since the condensate fraction depends only on
 the asymptotic properties of the one-particle Green function, and this 
 is given exactly via bosonization, we may conclude that the derived 
 result for this quantity is therefore, exact.
 The formula for the condensate fraction 
 may also be derived using other methods, most recently Liu\cite{Liu} has
 used the eigenfunctional theory to derive an expression for the same
 quantity that agrees with the expression derived in this article. 

 We apply this method to compute the velocity-velocity correlation function
 and demonstrate the power of this method in accounting for vortices
 in a gauge invariant manner.

\section{Density Phase Transformation}

Consider the action of nonrelativistic spinless bosons.
We use units such that $ \hbar = 2m = 1 $.
\begin{equation}
S_{free} = \int^{-i\beta}_{0} dt \mbox{        }
\int d^{d}x \mbox{          }
\psi^{\dagger}({\bf{x}})
\left( i \partial_{t} +  \nabla^{2}  \right)
\psi({\bf{x}})
\end{equation}
 In the path integral approach, $ \psi({\bf{x}}) $ is just a complex
 number that is defined at each point $ {\bf{x}} $. 
 Every complex number can be decomposed into a magnitude and a phase.
 Using the ideas explained in our earlier work\cite{Setlur}
\begin{equation}
\psi({\bf{x}}) = e^{-i \mbox{  }\Pi({\bf{x}}) }\mbox{        }
 \sqrt{ \rho({\bf{x}}) }
\end{equation}
From this we may write down the current,
\begin{equation}
{\bf{J}}({\bf{x}}) = -\rho({\bf{x}}) \mbox{         }
\nabla \Pi({\bf{x}}) 
\end{equation}
\begin{equation}
S_{free}  = \int^{-i\beta}_{0}dt \mbox{  }
\int d^{d}x \mbox{        }
\left[ \rho \mbox{      }\partial_{t} \mbox{  }\Pi
 - \rho  ( \nabla \Pi)^{2}
 - \frac{ (\nabla \rho)^{2} }{4 \rho} \right]
\end{equation}
 In deriving this action we have already used some boundary conditions.
 From the text by Kadanoff and Baym\cite{Kadanoff} we learn that the
 Green functions of bosons 
 obey the KMS (Kubo-Martin-Schwinger) boundary contitions.
 These boundary conditions
 translate in the path integral representation to,
\begin{equation}
\psi({\bf{x}},t-i\beta) = e^{\beta \mu} \mbox{        }
\psi({\bf{x}},t)
\end{equation}
 This in turn means that the number conserving product is invariant
 under this discrete time translation. To see this we examine,
 $ \psi^{\dagger}({\bf{x}},t-i\beta) 
 \psi({\bf{x}}^{'},t-i\beta) $. Observe that
 $  \psi^{\dagger}({\bf{x}},t-i\beta) =
 [\psi({\bf{x}},t+i\beta)]^{\dagger} = e^{-\beta \mu} \mbox{        }
\psi^{\dagger}({\bf{x}},t) $ and
 $ \psi({\bf{x}}^{'},t-i\beta) = e^{\beta \mu} \mbox{        }
\psi({\bf{x}}^{'},t) $. Thus,
 $ \psi^{\dagger}({\bf{x}},t-i\beta)  \psi({\bf{x}}^{'},t-i\beta) =  \psi^{\dagger}({\bf{x}},t)  \psi({\bf{x}}^{'},t) $.
 As pointed out in an earlier work\cite{Setlur} the phase variable
 may be written as a sum of two terms a position independent term
 which is the conjugate to the total number and a position dependent term
 that is related to currents and densitites. Thus,
 $ \Pi({\bf{x}}) = X_{0} + {\tilde{\Pi}}({\bf{x}}) $. We have just
 argued that $ {\tilde{\Pi}}({\bf{x}}) $ and $ \rho({\bf{x}}) $ 
 are invariant under the discrete time translation. Thus in order to
 preserve the KMS boundary condition we must impose
 $  X_{0}(t-i\beta) = X_{0}(t) + i \mbox{  }\beta \mu $.
 Therefore we find that the conjugate to the total number makes its
 presence felt in a very nontrivial manner.
 The boundary condition that has been used
 in deriving the action is $ N(-i\beta) = N(0) $ where
 $ N $ is the total
 number of particles.

\section{Gauge Transformations}

 Here we examine what sorts of changes are brought about by the imposition of
 local gauge invariance.
 Gauge transformations in the usual language is given by,
\begin{equation}
\psi^{'}({\bf{x}}) = e^{i \mbox{  }e\mbox{  }\Lambda} \mbox{   }
\psi({\bf{x}})
\end{equation}
\begin{equation}
A^{'}_{0} = A_{0} + \partial_{t} \Lambda
\end{equation}
\begin{equation}
{\bf{A}}^{'} = {\bf{A}} + \nabla \Lambda
\end{equation}
 In order to find an action invariant under
 these transformations,
 we replace derivatives by covariant derivatives(minimal coupling).
 Thus we have $ \partial_{\mu} \rightarrow D_{\mu}
 = \partial_{\mu} -i e \mbox{   }A_{\mu} $. In the usual language,
\begin{equation}
S  = \int 
\psi^{\dagger} 
 \left( i \partial_{t} + e \mbox{       } A_{0} 
 +  (\nabla - i \mbox{  }e\mbox{  }{\bf{A}})^{2} 
 \right)
\mbox{       }\psi
\end{equation}
 Here and henceforth by $ \int $ we mean
 $   \int^{-i\beta}_{0}dt \mbox{  }
\int d^{d}x \mbox{         } $
 Now we rewrite this as a sum of two parts the free term plus the
 term that couples to the gauge fields and finally the term involving
 only gauge fields namely, the curvature term.
\begin{equation}
S  =  S_{free} + S_{int} + S_{gauge}
\end{equation}
\begin{equation}
S_{free}  = \int \mbox{        }
\left[ \rho \mbox{      }\partial_{t} \mbox{  }\Pi
 - \rho  ( \nabla \Pi)^{2}
 - \frac{ (\nabla \rho)^{2} }{4 \rho} \right]
\end{equation}
\begin{equation}
S_{int} =  \int \mbox{         }
\rho\mbox{        } e \mbox{       } A_{0}
- e^{2} \mbox{    }\int \mbox{         }
\rho\mbox{           }{\bf{A}}^{2}
- 2e \mbox{    }\int \mbox{             }
\rho \mbox{        }\left( {\bf{A}} \cdot \nabla \Pi \right)
\end{equation}
\begin{equation}
S_{gauge} = - \frac{1}{4} \int \mbox{     }
 F^{2}
\end{equation}
This action is gauge invariant provided we have
\begin{equation}
\Pi^{'} = \Pi - e\mbox{  }\Lambda \mbox{                 };
 \mbox{             }
 \rho^{'} = \rho
\end{equation}

\section{The Jacobian Determinant}

In this section we attempt to rewrite the partition function in terms
of the density and phase variables. Consider the partition function 
 in the original Bose language.
\begin{equation}
Z = \int D[\psi] \mbox{         }D[\psi^{\dagger}]
 \mbox{             }e^{i S[\psi,\psi^{\dagger}] }
\end{equation}
In the density phase variable language we have,
\begin{equation}
Z = \int D[\rho] \mbox{         }D[\Pi] \mbox{         }
J(\Pi,\rho) \mbox{           }
 \mbox{             }e^{i S[\Pi,\rho] }
\end{equation}
 Here $ J(\Pi,\rho) $ is the appropriate Jacobian determinant that
 tells us how the measure transforms. Fortunately,
 this Jacobian is a constant.
 To see this we write the definition of $ J $ as,
\begin{equation}  
J(\Pi,\rho) =
 \left| \begin{array}{clcr} \frac{ \delta \psi }{ \delta \Pi } && 
 \frac{ \delta \psi }{ \delta \rho } \\
\frac{ \delta \psi^{\dagger} }{ \delta \Pi } && 
 \frac{ \delta \psi^{\dagger} }{ \delta \rho }
 \end{array}  \right| \mbox{          } = \mbox{          }-i
\end{equation}
 In our next article, we shall see that the density phase
 variable ansatz (DPVA) for fermions\cite{Setlur}
 which includes a phase functional that is nonlocal in position space 
  also leads to a Jacobian that is constant. 
 This can be checked using the Mathematica $ ^{TM} $ software. 

\section{ Propagator With Two-Body Forces }

 Here we compute the free propagator using path integrals in
 the density phase variable language. This is done to convince
 ourselves of the basic soundness of the approach. Also we 
 operate in the limit where the mean density is constant and
  so that we may ignore the density fluctuations in the long-wavelength
  limit.  This will be made precise soon.
 The main advantage of this approach is the ease with which we may treat
 interactions. Just to highlight this fact we also include
 density-density interactions.  
\begin{equation}
S_{free} \approx  \int \mbox{       }
\left[  \rho \mbox{       }
\partial_{t} \Pi - \rho_{0} (\nabla \Pi)^{2}
 - \frac{ (\nabla \rho)^{2} }{4 \rho_{0}}
\right] 
\end{equation}
Using the boundary conditions we may write,
\begin{equation}
\rho({\bf{x}},t) = \frac{1}{V}
 \sum_{ {\bf{q}}, n}e^{-i{\bf{q}}.{\bf{x}} }
\rho_{ {\bf{q}} n } \mbox{            }e^{-z_{n}t}
\end{equation}
\begin{equation}
\Pi({\bf{x}},t) = 
 -\mu \mbox{         }t 
 + \sum_{ {\bf{q}}, n}e^{i{\bf{q}}.{\bf{x}} }
X_{ {\bf{q}}n } \mbox{             }e^{z_{n}t}
\end{equation}
The action then may be written as follows. 
 We have added a two-body potential since we may
 do so without additional effort.
\[
S_{full} = i \mbox{   }\beta \mu \mbox{          }
\rho_{ {\bf{0}}, 0 } 
 + \sum_{ {\bf{q}}, n }
(-i\beta \mbox{  }z_{n}) \rho_{ {\bf{q}}n }X_{ {\bf{q}}n }
\]
\begin{equation}
+ i \beta N^{0} \sum_{ {\bf{q}}, n }
{\bf{q}}^{2} \mbox{       }X_{ {\bf{q}},n }
X_{ -{\bf{q}},-n }
+ \frac{i \beta }{4N^{0}} \sum_{ {\bf{q}}, n }{\bf{q}}^{2}
 \mbox{          }\rho_{ {\bf{q}}, n } \rho_{ -{\bf{q}}, -n }
+ i \beta \sum_{ {\bf{q}},n } \frac{ v_{ {\bf{q}} } }{2V} 
\rho_{ {\bf{q}}, n } \rho_{ -{\bf{q}}, -n }
\end{equation}
 Here $ z_{n} = 2\pi n/\beta $.
 From this we may write down a formal expression for the propagator.
 We set $ \rho \approx \rho_{0} $ in the argument of the square roots.
\begin{equation}
G({\bf{x}},t) = 
-i \rho_{0} \mbox{         }e^{i\mu t}
exp \left[ \frac{1}{4}
\sum_{ {\bf{q}},n }
\frac{(2 - e^{i{\bf{q}}.{\bf{x}} } e^{z_{n}t}
 - e^{-i{\bf{q}}.{\bf{x}} } e^{-z_{n}t})}
{ \beta N^{0} {\bf{q}}^{2} + \frac{1}{2}
\frac{ \beta^{2} z^{2}_{n} }{ \frac{\beta {\bf{q}}^{2} }{2N^{0}}
 + \frac{\beta v_{ {\bf{q}} } }{V} } }
 \right]
\end{equation}
 Here $ \omega^{2}_{ {\bf{q}} } = 
 {\bf{q}}^{2}({\bf{q}}^{2} + 2 \rho_{0}v_{ {\bf{q}} }) $
 is the Bogoliubov dispersion. 
 The Matsubara sums may be performed quite easily and we may 
  derive an expression for the
  equal-time version of the time-ordered propagator,
\begin{equation}
\gamma_{0} + S^{<}({\bf{x}},0)
 = -\frac{1}{4N^{0}}\sum_{ {\bf{q}} }
 \left( \frac{ ({\bf{q}}^{2} + 2 \rho_{0} v_{ {\bf{q}} })^{\frac{1}{2}} }
{ |{\bf{q}}| } - 1 \right)
(1- cos({\bf{q.x}}))
\end{equation}
where,
\begin{equation}
G^{<}({\bf{x}},0) = e^{ \gamma_{0} + S^{<}({\bf{x}},0) }
 \mbox{        }G^{<}_{0}({\bf{x}},0)
\end{equation}
 The main reason why the one-particle propagator
 is interesting is because while the
 Gaussian approximation leads to just Bogoliubov's theory as far as the
 the computation of density-density correlation functions are concerned,
 the one particle properties are singular in one dimension, 
 just as in the case of fermions in one dimension. 

\subsection{The Condensate Fraction at Zero Momentum}

 The analysis in the preceeding sections employs a Gaussian approximation
 which is valid if the density fluctuations are small compared to 
 the mean density. To see this more clearly we write the condition as follows,
\begin{equation}
\sqrt{ <\rho_{ {\bf{q}} } \rho_{ -{\bf{q}} } > } 
 \ll  N^{0}  
\end{equation}
where $ N^{0} $ is the total number of particles. 
The density-density correlation is given by
 $  <\rho_{ {\bf{q}} } \rho_{ -{\bf{q}} } > = N^{0} \mbox{    }S({\bf{q}}) $.
Using the same Gaussian approximation we may deduce that
\begin{equation}
S({\bf{q}}) = \frac{ |{\bf{q}}| }
{ ({\bf{q}}^{2}  + 2\rho_{0} v_{ {\bf{q}} })^{\frac{1}{2}} }
\end{equation}
Thus this scheme is self-consistent (as opposed to self-contradictory)
only if,
\begin{equation}
\frac{ |{\bf{q}}| }
{ ({\bf{q}}^{2}  + 2\rho_{0} v_{ {\bf{q}} })^{\frac{1}{2}} }
 \ll N^{0}
\end{equation}
If we assume that $ v_{ {\bf{q}} } = \lambda |{\bf{q}}|^{m+2} $ we have,
the condition( for $ N^{0} \gg 1 $ ),
\begin{equation}
(2\rho_{0} \lambda) \mbox{       }|{\bf{q}}|^{m} \gg -1
\end{equation}
 It would appear that so long as $ \lambda > 0 $ and $ m \leq -2 $,
 this holds for small $ |{\bf{q}}| $. Thus the delta-function 
 potential ( $ m = -2 $ ),
 the Coulomb potential in 2d ( $ m = -3 $ ) and 3d ( $ m = -4 $ )
 all obey the inequality for small $ |{\bf{q}}| $.
 For large enough $ |{\bf{q}}| $ the left hand side is zero 
 but $ 0 \gg -1 $ is not acceptable, hence the Gaussian approximation breaks
 down for large $ |{\bf{q}}| $. Thus it would appear that the results for 
 the correlation functions are exact in the asymptotic limit. 
 In real space, this is the $ |{\bf{x}}| \rightarrow \infty $ limit.
 Let us examine this limit of the one-particle Green function.
 To this end let us write a formal
 expression for the momenum distribution as
\begin{equation}
{\bar{n}}_{ {\bf{k}} } = f_{0} \mbox{       }N^{0} \mbox{        }
\delta_{ {\bf{k}}, 0 }
 + f_{ {\bf{k}} }
\end{equation}
 here $  f_{ {\bf{k}} } $ is a continuous
 function of $ |{\bf{k}}| $
 and is of order unity. $ 0 \leq f_{0} \leq 1 $ is the condensate fraction and 
 $ N^{0} $ is the total number of particles. The propagator in real space 
 is then given by,
\begin{equation}
< \psi^{\dagger}({\bf{x}},0) \psi({\bf{0}},0) > 
 = f_{0} \mbox{        }\rho_{0}
 + \frac{1}{V} \sum_{ {\bf{k}} } cos({\bf{k.x}}) f_{ {\bf{k}} }
\end{equation}
 In the ultra-asymptotic limit $ |{\bf{x}}| \rightarrow \infty $, 
 the cosine may be set equal to zero and we have just the first term. Thus
 it would appear that the Gaussian approximation being exact in this
 limit, yields the exact condensate fraction but not the exact
 $ f_{ {\bf{k}} } $. This is analogous to the assertion that in the Fermi case,
 the analogous method yields the exact quasiparticle-residue but not the
 full momentum distribution and in case the quasiparticle-residue is zero
 it yields the exact anomalous exponent. Here too when the condensate
 fraction is zero, we have to instead compute the anomalous exponent. 
 Thus the exact condensate fraction is given by,
\begin{equation}
f_{0} =
 exp\left[ -\frac{1}{4N^{0}}\sum_{ {\bf{q}} }
 \left( \frac{ ({\bf{q}}^{2} + 2 \rho_{0} v_{ {\bf{q}} })^{\frac{1}{2}} }
{ |{\bf{q}}| } - 1 \right)
\right]
\end{equation}
 For the delta-function interaction in 1d, the condensate fraction is zero.
 This could not have been guessed from Bogoliubov's theory that predicts
 (rather assumes) that a condensate always exists.
 In two dimensions, for the gauge potential
 $ v_{q} = const./q^2 $, the condensate fraction is zero. 
 We could now evaluate the total energy of the system (per particle)
 and compare with the results of Lieb and Liniger \cite{Lin} who
 solve their model in 1d or with the results of Schick\cite{Sch}
 which is valid in 2d but we do not expect the comparisons to be favorable
 since for the total energy to come out right we need the
 exact $ f_{ {\bf{k}} } $ rather than $ f_{0} $ which does not contribute
 at all. Since the Gaussian approximation does not give us this we
 shall not bother performing this calculation. 

\subsection{Anomalous Exponents}

 In case the condensate fraction is zero, we have the bosonic analog of
 the Luttinger liquid or the Lieb-Liniger liquid. Here we have to
 instead compute the anomalous exponents which are also given exactly 
 in the Gaussian approximation. Interestingly we may address the 
 question whether the condensate is destroyed in more than one dimension
 also. In passing we note that the main purpose of this article is
 to hint at the usefullness of this approach in studying {\it{fermions}}
 coupled to gauge fields. There we have to polar decompose Grassmann variables
  - an exercise which is still in its infancy, although considerbale
 inroads have been made by the author.
 For a delta-function interaction in 1d it is well-known 
 that there is no condensation and instead we have to compute the anomalous
 exponent of the propagator. It is clear from the preceeding sections that
 we may write, 
\begin{equation}
\gamma_{0} + S^{<}(x,0) 
 \approx -\frac{1}{4 \pi \rho_{0} }  \int_{0}^{\infty} dq 
\mbox{      }
 \frac{  (2 v_{ 0 }\rho_{0})^{\frac{1}{2}} }
{ |q| }
(1- cos(q.x))
 \sim - \gamma Ln(|x|)
\end{equation}
where $ \gamma = (2 v_{ 0 }\rho_{0})^{ \frac{1}{2} }/(4 \pi \rho_{0}) $
 is the anomalous exponent. 
 The interesting question is whether
 we can have the destruction of the condensate in two or three dimensions
 for realistic potentials ? This would be the bosonic analog of the question 
 `Does Fermi liquid theory break down in two or three dimensions ?' 
 In two dimensions, for the interaction $ v_{q} = const./q^2 $ 
 (gauge potential) the condensate fraction is zero. It would appear that 
 for realistic Fourier transformable potentials in three dimensions
 we have Bose condensation but not in less than three dimensions.
 We have derived an exact formula for the condensate fraction that is valid for
 functional forms of the interaction that are Fourier transformable and 
 are subject to the constraint mentioned previously namely that the
 long-wavelength limit is exactly given by the Gaussian approximation.  

\section{Correlation Functions With Gauge Fields}

 When one is performing the path integral
 with gauge fields, one must be careful about preserving
 gauge symmetry.
 This is the crucial aspect that may be elegantly treated in
 the present approach. The usual method for treating such constraints 
 in the context of path integrals
 is the well-known Faddeev-Popov method
 \cite{Zinn}. However we are able
 to treat gauge symmetry in all its generality
 without ever having to fix the gauge at any time.
 Thus the path integral is to be performed such that the following
\begin{equation}
e\mbox{   }A^{'}_{0} + \partial_{t} \Pi^{'} 
 = e\mbox{   }A_{0} + \partial_{t} \Pi
\end{equation}
\begin{equation}
 e\mbox{    } {\bf{A}}^{'} + \nabla \Pi^{'} 
 =  e\mbox{    }{\bf{A}} + \nabla \Pi
\label{EQBI}
\end{equation}
 constraints are obeyed. We may therefore simply solve for the
 gauge fields straight away in terms of the conjugate variables.
 Define an arbitrary gauge constant $ C $.
 This is constant in the sense that
 changes in the phase of the field cancel out the changes in the vector
 potential. However, in the end we will have to integrate over this variable
 as well. It will be shown that in the case when $ e = 0 $ we recover the
 noninteracting propagator. 
\begin{equation}
C_{0} =  e \mbox{       }A_{0} + \partial_{t} \Pi
\end{equation}
\begin{equation}
{\bf{C}} = e \mbox{      } {\bf{A}} + \nabla \Pi
\label{EQNG}
\end{equation}
  In the noninteracting limit ($ e \equiv 0 $), $ \Pi^{'} $ and $ \Pi $ 
  differ at most by a trivial constant and
  hence the vector $ {\bf{C}} $ is irrotational.
 This observation will be made use of later.
 The action now may be recast in terms of the gauge constant
 and the density and phase variables. 
 The field tensor is then simply given by,
   $ F_{\mu\nu} = \partial_{\mu} C_{\nu} - \partial_{\nu} C_{\mu} $.
 Thus the total action simply reads,
\begin{equation}
S = - \int \frac{ (\nabla \rho)^{2} }{4 \rho} 
 + \int \rho  \mbox{          }C_{0}
- \int \rho \mbox{           }{\bf{C}}^{2}
 - \frac{1}{4e^{2}} \int F^{2}
\label{SNEW}
\end{equation}
 We may see by examining Eq.(~\ref{SNEW}) 
 that in the limit $ e \rightarrow 0 $, the field tensor has to be zero
 for the partition function to be nonvanishing, thus in this limit we must have
 $ C_{\mu} = \partial_{\mu} \Pi $ for some scalar $ \Pi $ so that
 $ F_{\mu \nu} \equiv 0 $. This then gives us the action functional of 
 the free theory as it should. 

 The next question that is worth answering is the following. How should the
 current operator be defined when there are gauge fields ? In particular 
 is the current operator gauge invariant ? The density surely is
 gauge invariant. If 
 we would like to treat currents and densities as part of a canonical set
 of variables then we better have a definition that is gauge invariant.
 The definition $ {\bf{J}} = -\rho\nabla \Pi $ is not gauge
 invariant. Indeed, it changes by additive amounts everytime a gauge
 transformation is performed. In order to remedy this we redefine the current
 to be $  {\bf{J}} = -\rho\mbox{       }{\bf{C}} $. As we have seen,
 in the noninteracting limit, this reduces to the form already shown,
 since there $ {\bf{C}} = \nabla \Pi $.  In general however, it does not. 
 Now we would like an expression for the field operator also in terms
 of $ C $. Using the DPVA\cite{Setlur} in terms of currents and densities
 we may write,
\begin{equation}
\psi({\bf{x}}) = e^{-iX_{0} + i \int^{ {\bf{x}} }d{\bf{l}} \cdot
\frac{ {\bf{J}} }{\rho} } \sqrt{ \rho({\bf{x}}) }
\end{equation}
Using the formula for the current in terms of the gauge constant we may write,
\begin{equation}
\psi({\bf{x}},t) = e^{-iX_{0}(t)
 - i \int^{ {\bf{x}} }d{\bf{l}} \cdot
 {\bf{C}}({\bf{x}}^{'},t)  } \sqrt{ \rho({\bf{x}},t) }
\end{equation}
Computing the propagator of the interacting theory 
 then means evaluating the following path integral.
\begin{equation}
G({\bf{x}},t) = \frac{-i}{Z}
\int D[C] D[\rho] \mbox{         } e^{i \mbox{  }S} \mbox{   }
\psi({\bf{x}},t) 
\psi^{\dagger}({\bf{0}},0) 
\end{equation}
and
\begin{equation}
Z = \int D[C] D[\rho] \mbox{         } e^{i \mbox{  }S} \mbox{   }
\label{EQNZ}
\end{equation}
 where $ S $ is given as in Eq.(~\ref{SNEW}).
 The integration over the gauge constant $ C $ may be justified as follows.
 In the original path integral we had to integrate over
 $ \psi $, $ \psi^{\dagger} $ and the four $ A^{\mu} $ 's. 
 Thus we had six variables in all, except that the gauge constraint meant
 that one of the variables, say the phase of the field depended on 
 the gauge fields thus the number of independent variables reduced to five.
 In the definition Eq.(~\ref{EQNZ}) we have a similar situation -  
 integrating over the four components of the gauge constant plus the
 density $ \rho $ makes five variables as it should. 
 The action of the free gauge field is given by,
\begin{equation}
S_{guage} = \frac{1}{2e^{2}} \int
 ( \partial_{0} {\bf{C}} - \nabla C_{0})^{2} 
 + \frac{v^{2}_{l}}{2e^{2}} \int {\bf{C}} \mbox{   }\nabla^{2} {\bf{C}} 
 + \frac{v^{2}_{l}}{2e^{2}} \int ( \nabla \cdot {\bf{C}} )^{2}
\end{equation}
 A word regarding units is appropriate. We have set $ \hbar = 2m = 1 $.
 The presence of gauge fields implies that there
 is a further dimensionful parameter namely the speed
 of light which we denote by $ v_{l} $.  If we set $ v_{l} = 1 $ then all 
 quantities are dimensionless. This is undesirable. Henece we retain the
 speed of light as it is. The various quantites in this work now have the
 following dimensions. $ [\rho_{ {\bf{k}} n }] = 1 $ but
 $ [\rho({\bf{x}},t)] = [V]^{-1} $ ;
$ [{\bf{C}}({\bf{x}},t)] = [{\bf{C}}_{ {\bf{k}} n }] = 
[{\bf{k}}] = [\nabla] = [v_{l}] $ ;
 $ {\bf{k}}^{2} = [Energy] = [\partial_{0}] $ ;
 $ \beta = [Energy]^{-1} $ and finally
 $ [e^{2}/V] = [Energy]^{2} $.   
Using the decomposition into the various modes we may write,
\[
S = \frac{ i\beta  }{ 4N^{0} }\sum_{ {\bf{k}} }{\bf{k}}^{2} 
 \mbox{  }\rho_{ {\bf{k}}, n } \rho_{ -{\bf{k}}, -n }
 + (-i\beta) \sum_{ {\bf{k}}, n } \rho_{ {\bf{k}},n }
C^{0}_{ {\bf{k}},n } 
 + (i\beta N^{0}) \sum_{ {\bf{k}}, n }
 {\bf{C}}_{ {\bf{k}}, n } \cdot {\bf{C}}_{ -{\bf{k}}, -n }
\]
\[
+(i\beta) \frac{V}{2e^{2}} \sum_{ {\bf{k}}, n }
 z^{2}_{n} {\bf{C}}_{ {\bf{k}},n } \cdot
 {\bf{C}}_{ -{\bf{k}}, -n } 
- (i\beta) \frac{V}{2e^{2}} \sum_{ {\bf{k}}, n } 
(2i z_{n} \mbox{        }{\bf{k}})
{\bf{C}}_{ {\bf{k}},n } C^{0}_{ -{\bf{k}}, -n }
 - (i\beta) \frac{V}{2e^{2}} 
\sum_{ {\bf{k}} } {\bf{k}}^{2} 
 C^{0}_{ {\bf{k}},n }C^{0}_{ -{\bf{k}}, -n }
\]
\begin{equation}
 + (i\beta) \mbox{        }v^{2}_{l} \mbox{     }\frac{V}{2e^{2}}
\sum_{ {\bf{k}}, n } 
{\bf{k}}^{2} \mbox{  }{\bf{C}}_{ {\bf{k}}, n }
 \cdot {\bf{C}}_{ -{\bf{k}}, -n }
 - (i\beta) \mbox{       }v^{2}_{l} \mbox{       }\frac{V}{2e^{2}}
\sum_{ {\bf{k}}, n } 
({\bf{k}} \cdot {\bf{C}}_{ {\bf{k}}, n })
({\bf{k}} \cdot {\bf{C}}_{ -{\bf{k}}, -n })
\end{equation}
After integrating out the $ C^{0} $ we have,
\[
S = \frac{i\beta}{4N^{0}}\sum_{ {\bf{k}}n }
\left( {\bf{k}}^{2} + \frac{2 \rho_{0} e^{2} }{ {\bf{k}}^{2} } \right)
\rho_{ {\bf{k}}, n} \rho_{ -{\bf{k}}, -n }
+ (i\beta) \sum_{ {\bf{k}}, n } \frac{ iz_{n} }{ {\bf{k}}^{2} }
 \rho_{ {\bf{k}} n } ({\bf{k}} \cdot {\bf{C}}_{ {\bf{k}} n })
\]
\[
+ (i\beta) \sum_{ {\bf{k}} n }
\left( N^{0} + v^{2}_{l}\frac{V}{2e^{2}} {\bf{k}}^{2}
 \right)  {\bf{C}}_{ {\bf{k}}, n } \cdot {\bf{C}}_{ -{\bf{k}}, -n } 
 - (i\beta) v^{2}_{l} \mbox{  }\frac{V}{2e^{2}} 
\sum_{ {\bf{k}} n } ({\bf{k}} \cdot {\bf{C}}_{ {\bf{k}} n })
\mbox{        } ({\bf{k}} \cdot {\bf{C}}_{ -{\bf{k}}, -n })
\]
\begin{equation}
+ (i\beta) \frac{V}{2e^{2}} \sum_{ {\bf{k}}n }
z^{2}_{n} \mbox{   }{\bf{C}}_{ {\bf{k}} n } \cdot {\bf{C}}_{ -{\bf{k}},-n }
- (i\beta) \frac{V}{2e^{2}} \sum_{ {\bf{k}}n }
\frac{ z^{2}_{n} }{ {\bf{k}}^{2} }
\mbox{   }( {\bf{k}} \cdot {\bf{C}}_{ {\bf{k}} n } )
( {\bf{k}} \cdot  {\bf{C}}_{ -{\bf{k}},-n })
\end{equation}
 Observe that in the limit $ e \rightarrow 0 $, we must have
 $ {\bf{C}}_{ {\bf{k}}, n } = {\bf{k}} \mbox{      }X_{ {\bf{k}}, n } $ 
 for the partition function to be nonvanishing. Thus we recover
 the noninteracting theory in this limit. 
 Now we would like to compute two quantities. One is the dynamical
 density-density correlation function. Here we would like to see 
 radiation corrections to corresponding spectral function. However, it is 
 not present at the Gaussian level at which we are presently operating.
 Thus we shall be content at reproducing results 
 equivalent to Bogoliubov theory. 
 This is a novel way of recovering the Bogoliubov spectrum by introducing
 gauge fields and using the path integral formalism.  The other quantity
 of interest is the circulation of the gauge constant. This quantity is
 zero in the nonintearcting case since the vector $ {\bf{C}} $ may be
 expressed as the gradient of a scalar.
 In fact we may see from the expression of the current that
 the velocity of the fluid is simply given by
 $ {\bf{v}} = -{\bf{C}} $. This is irrotational
 in the absence of gauge fields.
 In general however this is not
 the case. The circulation of the velocity
 around a closed loop $ P $ is given by,
 $ V(P,t) = -\oint_{ P } d {\bf{l}} \cdot {\bf{C}}({\bf{x}},t) $.
 Physically, the quantity $ V(P,t)  $ is a measure of the strength of
 vortices in the system.
 The average $ < V(P,t) > = 0 $ is trivially zero. Thus we have to
 examine the fluctuation $ < V^{2}(P,t) >^{\frac{1}{2}} $.
 First the density-density correlation function.
 In order to evaluate this
 we must first perform the integral
 :  $ \int D[{\bf{C}}] \mbox{       }e^{iS} $.
 Again here we may use the usual procedure of translating the
 integration variable by a constant amount and we obtain the following
 effective action. 
\begin{equation}
S = \frac{ i \beta }{4 N^{0}} \sum_{ {\bf{k}}, n }
\left( {\bf{k}}^{2} + \frac{ z^{2}_{n} + 2 e^{2} \rho_{0} }{ {\bf{k}}^{2} }
 \right) \rho_{ {\bf{k}}, n } \rho_{ -{\bf{k}}, -n }
\label{EQNAC}
\end{equation}
From this we may immediately deduce the static structure factor as
\begin{equation}
S({\bf{k}}) = \frac{ {\bf{k}}^{2} }{ \omega_{ {\bf{k}} } }  
\end{equation}
where the Bogoliubov dispersion $ \omega_{ {\bf{k}} } $ is given by
 the positive real solution to,
\begin{equation}
 {\bf{k}}^{2} + \frac{ (iz)^{2} + 2 e^{2} \rho_{0} }{ {\bf{k}}^{2} }
 = 0
\end{equation}
In other words,
\begin{equation}
\omega_{ {\bf{k}} } = |{\bf{k}}|
 \sqrt{ {\bf{k}}^{2} + 2 \rho_{0} v_{ {\bf{k}} } }
\end{equation}
 where $ v_{ {\bf{k}} } = e^{2}/ {\bf{k}}^{2} $. Therefore
 as is well-known, independent of
 the dimensionality of space, the interaction in the presence of gauge
 fields is forced to be of the form $ \sim 1/{\bf{k}}^{2} $.
 In three space dimensions, this corresponds to Coulomb interaction.
 This is a necessary consistency check.

 Some may want to evaluate the propagator. For this we have to 
 evaluate the correlation function
 $ < {\bf{C}}^{i}({\bf{x}},t)  {\bf{C}}^{j}({\bf{0}},0) > $.
 There is a more interesting reason to study this, namely to compute
 the vortex strength. We have found that even though at the Gaussian
 level the density-density correlation is unremarkable,
 the velocity-velocity correlation function does exhibit some new physics.
 To see this we note that integrating out the velocity variable 
 means replacing
 $ {\bf{C}}_{ {\bf{k}} n } \rightarrow 
 {\bf{C}}_{ {\bf{k}} n } + {\bf{k}}  \Lambda_{ {\bf{k}} n } $
 for an appropriate $ \Lambda $. This makes all the terms involving 
 the speed of light and most other terms drop out and we are led to  
 Eq.(~\ref{EQNAC}). However if we integrate out the density first
 and retain the velocity as it is then we find that the final
 action is no longer so simple. In particular, it will involve both
 longitudinal and transverse terms, the latter being responsible for
 vortices as we shall see. If we integrate out the $ \rho $ first we are
 led to the following effective action. Here we have also
 introduced a source for velocity. 
\[
S_{0} = -i \beta N^{0} \sum_{ {\bf{k}} n }
\frac{ (iz_{n}/{\bf{k}}^{2})^{2} }
{ \left( {\bf{k}}^{2} + 2\rho^{0} e^{2}/{\bf{k}}^{2} \right) }
({\bf{k}} \cdot {\bf{C}}_{ {\bf{k}} n })
({\bf{k}} \cdot {\bf{C}}_{ -{\bf{k}},-n })
 + \sum_{ {\bf{k}} n } {\bf{A}}_{ {\bf{k}} n } \cdot {\bf{C}}_{ {\bf{k}} n }
\]
\[
+ (i\beta) \sum_{ {\bf{k}} n }
\left( N^{0} + v^{2}_{l}\frac{V}{2e^{2}} {\bf{k}}^{2}
 \right)  {\bf{C}}_{ {\bf{k}}, n } \cdot {\bf{C}}_{ -{\bf{k}}, -n } 
 - (i\beta) v^{2}_{l} \mbox{  }\frac{V}{2e^{2}} 
\sum_{ {\bf{k}} n } ({\bf{k}} \cdot {\bf{C}}_{ {\bf{k}} n })
\mbox{        } ({\bf{k}} \cdot {\bf{C}}_{ -{\bf{k}}, -n })
\]
\begin{equation}
+ (i\beta) \frac{V}{2e^{2}} \sum_{ {\bf{k}}n }
z^{2}_{n} \mbox{   }{\bf{C}}_{ {\bf{k}} n } \cdot {\bf{C}}_{ -{\bf{k}},-n }
- (i\beta) \frac{V}{2e^{2}} \sum_{ {\bf{k}}n }
\frac{ z^{2}_{n} }{ {\bf{k}}^{2} }
\mbox{   }( {\bf{k}} \cdot {\bf{C}}_{ {\bf{k}} n } )
( {\bf{k}} \cdot  {\bf{C}}_{ -{\bf{k}},-n })
\label{EQNS0}
\end{equation}
 It can bee seen from Eq.(~\ref{EQNS0}) that in the
 limit $ v_{l} \rightarrow \infty $, the partition function
 is nonvanishing only if
 $ {\bf{k}}^{2} {\bf{C}}_{ {\bf{k}} n } \cdot {\bf{C}}_{ -{\bf{k}},-n }
 = ({\bf{k}} \cdot  {\bf{C}}_{ {\bf{k}} n }) ({\bf{k}} \cdot
 {\bf{C}}_{ -{\bf{k}}, -n }) $. In other words, only if
 $ {\bf{C}}_{ {\bf{k}} n } = {\bf{k}} \Lambda_{ {\bf{k}} n } $.
 This is the same as saying that the velocity is the gradient
 of some scalar $ {\bf{v}} = -\nabla \Lambda $. Thus in this limit
 there are no vortices. Vortices arise as a result of a finite speed of 
 light, in other words due to radiation corrections. 
 Now we evaluate the partition function
 $ {\tilde{Z}}([{\bf{A}}]) =  Z([{\bf{A}}])/ Z([{\bf{0}}]) $ where
 $ Z([{\bf{A}}]) = \int D[{\bf{C}}] \mbox{    }e^{iS_{0}} $.
 After some tedious algebra we arrive at,
\begin{equation}
{\tilde{Z}}({\bf{A}}) = exp \left[ -
\frac{1}{2}\sum_{ {\bf{k}} n } f({\bf{k}}, n ) 
({\bf{k}} \cdot {\bf{A}}_{ {\bf{k}}, n })
({\bf{k}} \cdot {\bf{A}}_{ -{\bf{k}}, -n })
 - \frac{1}{2}\sum_{ {\bf{k}} n } g({\bf{k}},n) 
 {\bf{A}}_{ {\bf{k}}, n } \cdot {\bf{A}}_{ -{\bf{k}}, -n } \right]
\end{equation}
where,
\[
f({\bf{k}},n) = (2\beta N^{0})^{-1}
\left[ \frac{ (iz_{n}/{\bf{k}}^{2})^{2} }
{ {\bf{k}}^{2} + 2 \rho^{0} e^{2}/{\bf{k}}^{2} } 
 + \frac{1}{ 2 \rho^{0} e^{2} }
(v^{2}_{l} + \frac{ z^{2}_{n} }{ {\bf{k}}^{2} }) \right]
\]
\begin{equation}
\times \left( 1 + \frac{1}{ 2 \rho^{0} e^{2} }
(v^{2}_{l} {\bf{k}}^{2} + z^{2}_{n}) \right)^{-1}
\left( 1 + \frac{ z^{2}_{n}/{\bf{k}}^{2} }
{ {\bf{k}}^{2} + 2 \rho^{0}e^{2}/{\bf{k}}^{2} } \right)^{-1}
\end{equation}
\begin{equation}
 g({\bf{k}},n) = (2 \beta N^{0})^{-1}
\left( 1 + \frac{1}{ 2 \rho^{0} e^{2} }
(v^{2}_{l} {\bf{k}}^{2} + z^{2}_{n}) \right)^{-1}
\end{equation}
 We may deduce a formula for the time ordered velocity-velocity
 correlation function,
\begin{equation}
\left< T \mbox{       }{\bf{C}}^{i}({\bf{x}}^{'},t^{'})
 \mbox{   } {\bf{C}}^{j}({\bf{x}},t)  \right>
 = \left[ \nabla^{i}_{ {\bf{x}}^{'} } \nabla^{j}_{ {\bf{x}} } \mbox{       }
 F({\bf{x}}^{'}-{\bf{x}},t^{'}-t) 
 +  H({\bf{x}}^{'}-{\bf{x}},t^{'}-t)  \mbox{    }\delta_{i,j} \right]
\end{equation}
\begin{equation}
F({\bf{x}}^{'},t^{'}) = \sum_{ {\bf{k}}, n }
f({\bf{k}},n) \mbox{      }e^{ i{\bf{k}}.{\bf{x}}^{'} }
e^{-z_{n} t^{'}} \mbox{             };
\mbox{                 }
H({\bf{x}}^{'},t^{'}) = \sum_{ {\bf{k}}, n }
g({\bf{k}},n) \mbox{      }e^{ i{\bf{k}}.{\bf{x}}^{'} }
e^{ -z_{n}t^{'} }
\end{equation}
 The $ \Sigma $ and $ \sigma $ below are defined in the appendix.
\begin{equation}
\Sigma(t-t^{'};P,P^{'}) = -i\mbox{      }
\oint_{ {\bf{l}}^{'} \in P^{'} } 
\oint_{ {\bf{l}} \in P }  \mbox{         }
 d{\bf{l}}^{'} \cdot d{\bf{l}} \mbox{         }
  H({\bf{x}}^{'}-{\bf{x}},t^{'}-t)
\end{equation}
 Also we may write for the propensity to create
 vortices(defined in the appendix),
\begin{equation}
\sigma(P,P^{'}) =  -\frac{ e^{2} }{ 4 \pi v^{2}_{l} }\mbox{       }
\oint_{ {\bf{l}}^{'} \in P^{'} } 
\oint_{ {\bf{l}} \in P }  \mbox{         }
 d{\bf{l}}^{'} \cdot d{\bf{l}} \mbox{         }
 \frac{ e^{ -\lambda |{\bf{x}}^{'}-{\bf{x}}| } }
{  |{\bf{x}}^{'}-{\bf{x}}| }
\end{equation}
 where $ \lambda =  (2 \rho^{0} e^{2}/v^{2}_{l})^{\frac{1}{2}} $.
 This quantity looks finite. Hence there is no vortex instability
 in the Gaussian approximation. Let now compute the vortex strength as
 defined earlier.
\begin{equation}
\left< V^{2}(P,t) \right> = \oint_{ {\bf{l}}^{'} \in P } 
\oint_{ {\bf{l}} \in P }  \mbox{         }
 d{\bf{l}}^{'} \cdot d{\bf{l}} \mbox{         }
H({\bf{x}}^{'}-{\bf{x}},0)
\end{equation}
\begin{equation}
H({\bf{x}}^{'}-{\bf{x}},0) = \frac{ e^{2} }{ 2V }
\sum_{ {\bf{k}} }
\frac{ e^{ i{\bf{k}}.({\bf{x}}^{'}-{\bf{x}}) } }
{ \omega_{l}( {\bf{k}} ) } 
\mbox{    }
coth( \frac{ \beta \omega_{l}( {\bf{k}} ) }{2} )
\end{equation}
where $ \omega_{l}( {\bf{k}} ) =
 (2 \rho^{0} e^{2} + v^{2}_{l} {\bf{k}}^{2})^{\frac{1}{2}} $.
 This dispersion shows that the photons have acquired mass by coupling
 to matter. 
 An explicit evaluation of this and further analysis is possible and
 will not be done here since our interest is merely to highlight the usefulness
 of these ideas. However some intutive justification of the above
 formulas is in order. First the vortex strength and
  suceptibility are zero for a noninteracting system.
 This is hardly surprising since the velocity operator of free bosons is
 irrotational in our formalism. The author has tried to prove this
 independently using the second quantization
 formalism but has been unsuccesful
 in proving it rigorously. It appears that the only obstruction to this 
 conclusion at least at the formal mathematical level, is the lack of
 self-adjointness of the canonical conjugate to the total number of bosons
 \cite{Setlur} \cite{Setlur3}. In the high density limit this is not
 a problem. Thus as far as the asymptotics are concerned, our approach 
 provides reliable answers. In other words, the expression
 for the velocity-velocity correlation function is valid in the large
 seperation $ |{\bf{x}}-{\bf{x}}^{'}| \rightarrow \infty $ limit.
 However the main message is that the vortices are due to the finiteness of
 the speed of light.
 That is, even in an interacting system, there are
 no vortices if we include only the longitudinal density-density interactions.
 Coupling to transverse radiation fields is needed to generate vortices.
 This may be understood by realizing that in order to create vortices we 
 have to supply angular momentum. Photons being spin-1 particles, possess the
 angular momentum that the charged bosons use to generate vortices.

\section{Conclusions}

 We have written down formulas for the one-particle Green function
 of a homogeneous Bose system that is exact in the asymptotic limit.
 From the exact asymptotic form
 we may extract an exact formula for the condensate
 fraction and if it is zero an exact formula for the anomalous exponent. 
 We have also computed the vortex strength and shown that radiation corrections
 are responsible for the vortices. 
 No other quantity is given exactly in the formalism outlined. The
 total energy per particle, roton minimum and other important physical
 attributes will have to involve going beyond the Gaussian approximation.
 Nevertheless, the agenda in the immediate future 
  is to apply these ideas to fermions and then
 write down a theory of neutral matter with nuclei and electrons 
 treated on an equal footing. 

\section{Acknowledgements}

 I would like to thank Prof. G. Baskaran and Prof. G. Menon for their 
 input and useful suggestions. Prof. G. Menon has provided many of 
 the references to well-known literature on Bose systems.

\section{ Appendix }

In this section, we derive a Kubo-like formula that relates the 
propensity for vortices emerging ( an analog of d.c. conductivity )
with microscopic velocity-velocity correlation functions.
Let us write down the following formula for the vortex strength
with a source for vortices in the interaction representation.
As defined in the main text, the vorex strength may be written as follows.
\begin{equation}
 V(P,t) = - \oint_{P} d{\bf{l}} \mbox{       } \cdot
 \mbox{       }{\bf{C}}({\bf{x}},t) 
\end{equation}
\begin{equation}
\left< V(P,t) \right> = \frac{ \left< T \mbox{      }S \mbox{         }
{\hat{V}}(P,t) \right> }{ \left< T \mbox{      }S \right> }
\end{equation}
\begin{equation}
S = exp \left[ -i \int^{-i\beta}_{0} dt^{'} \mbox{          }
\sum_{P^{'}} {\hat{V}}(P^{'},t^{'}) W(P^{'},t^{'}) \right] 
\end{equation}
Here $ W(P,t) $ is a whirlpool source for the vortex $ V(P,t) $.
From linear response theory, we may expect that for weak sources
the vortex strength is proportional to the convolution of the
 source with a linear response coefficient.
\begin{equation}
\left< V(P,t) \right> = \sum_{P^{'}} \int^{-i\beta}_{0} dt^{'}
\Sigma(t-t^{'};P,P^{'})W(P^{'},t^{'}) 
\end{equation}
From this we see that,
\begin{equation}
\Sigma(t-t^{'};P,P^{'}) = \left( \frac{ \delta }
{ \delta W(P^{'},t^{'}) } \left< V(P,t) \right> \right)_{ W \equiv 0 }
\end{equation}
\begin{equation}
\Sigma(t-t^{'};P,P^{'}) = -i\mbox{      }
 \left< T \mbox{        }{\hat{V}}(P^{'},t^{'})
{\hat{V}}(P,t) \right> 
\end{equation}
If the whirlpool source is time-independent, we may write,
\begin{equation}
\left< V(P) \right> = \sum_{P^{'}} 
\sigma(P,P^{'})W(P^{'}) 
\end{equation}
\begin{equation}
\sigma(P,P^{'}) = \int^{-i\beta}_{0} dt^{'}
\Sigma(t-t^{'};P,P^{'})
\end{equation}
 If $ \sigma(P,P^{'}) = \infty $ then it signals instability
 to vortices, namely arbitrarily weak pertubations can lead to vortices.
 Thus we have,
\begin{equation}
\Sigma(t-t^{'};P,P^{'}) = -i\mbox{      }
\oint_{ P^{'} }  d{\bf{l}}_{i}^{'} \mbox{      } 
\mbox{        }\oint_{P}  d{\bf{l}}_{j}  \mbox{         }
 \left< T \mbox{        }{\bf{C}}^{i}({\bf{x}}^{'},t^{'})
 {\bf{C}}^{j}({\bf{x}},t) \right> 
\end{equation}

\end{document}